\documentclass[]{aastex631}

\def\eqref#1{(\ref{eq:#1})}

\begin{document}

\title{Impact of Climate States and Seasons on Future Exo-Earth Observations}

\author{Kyle Batra} 
\affiliation{Department of Earth, Atmospheric, and Planetary Science, Purdue University, West Lafayette, IN, USA}
\affiliation{NASA Network for Ocean Worlds Exo-oceanography Team}
\affiliation{Alternative Earths NASA ICAR Team}
\author{Stephanie Olson}
\affiliation{Department of Earth, Atmospheric, and Planetary Science, Purdue University, West Lafayette, IN, USA}
\affiliation{NASA Network for Ocean Worlds Exo-oceanography Team}
\affiliation{Alternative Earths NASA ICAR Team}
\author{Vincent Kofman}
\affiliation{Centre for Planetary Habitability (PHAB), Department of Geoscience. Oslo University, Oslo, Norway}

\begin{abstract}
Many planetary parameters impact the climate state of Earth-like exoplanets and could vary significantly from those on Earth. However, some of these parameters may be impossible to observe, causing ambiguity in determining exoplanet climate and characterizing their atmospheric features. We explore how distinct planetary climate states impact their reflectance spectra to reduce uncertainty in the interpretation of future direct imaging observations, such as with the Habitable Worlds Observatory. We find that worlds with the same atmospheric composition but distinct climate states have notable differences in apparent albedos and feature detectability. An additional consequence is that the exposure time required to detect atmospheric features and biosignatures, such as O$_2$, will depend on climate state, with icier worlds being more favorable for biosignature detection while ice-limited worlds may be more habitable. We find that clouds improve the strength and detectability of atmospheric features in reflected light, especially for ice-limited low albedo worlds. We find temporal variation in the strength of spectra at different seasons on high obliquity worlds, causing the required time to resolve atmospheric features to vary between the equinoxes and solstices. This abiogenic seasonality could be detectable through repeated direct imaging observations and may help inform the planetary climate state, especially in combination with constraints on inclination and mass. Our work elevates the importance of astrometry performed concurrently with direct imaging for characterizing climate state and planetary habitability of exoplanets. Interpretation of future spectroscopic observations must also account for temporal variations created by obliquity when searching for biosignatures.
\end{abstract}


\section{Introduction} \label{sec:intro}

Earth-like habitable zone planets can have a variety of climate states, ranging from ice-free (no surface ice) to globally glaciated (oceans covered in surface ice). The climate state is affected by several planetary parameters that may vary significantly from those on Earth \citep{kane_obliquity_2017}. The ice-albedo positive feedback also amplifies minor changes in radiative forcing which impacts climate, introducing hysteresis and bistability on G-star (Sun-like) worlds \citep{curry_sea_1995, shields_effect_2013}. Past work has shown that ocean salinity can have a major influence on climate and habitability of G-star worlds in the habitable zone by modulating sea ice extent through its control of the freezing point of seawater \citep{del_genio_habitable_2019} as well as ocean circulation and heat transport to the poles \citep{cullum_importance_2016, cael_oceans_2017, olson_effect_2022, batra_climatic_2024}. Orbital parameters, such as planetary obliquity, can also impact climate state by controlling the spatial and temporal patterns of stellar flux across a planet latitudinally \citep{dressing_habitable_2010, armstrong_effects_2014, colose_enhanced_2019, jernigan_superhabitability_2023}. However, ocean salinity is impossible to observe on exoplanets while obliquity will be very challenging to remotely constrain \citep{schwartz_inferring_2016, lustig-yaeger_detecting_2018, adams_signatures_2019}, which can introduce uncertainty in the interpretation of future exoplanet observations \citep{olson_oceanographic_2020, batra_climatic_2024, batra_synergistic_2026}. Thus, knowing the instellation and the abundance of greenhouse gases like CO$_{2}$ is insufficient to reliably characterize exoplanet climate and assess habitability.

Characterizing terrestrial exoplanets orbiting within the habitable zone of G-stars with a direct imaging telescope is a major goal in the Astro2020 Decadal Survey \citep[e.g.,][]{committee_on_exoplanet_science_strategy_exoplanet_2018, committee_on_the_planetary_science_and_astrobiology_decadal_survey_origins_2022, committee_for_a_decadal_survey_on_astronomy_and_astrophysics_2020_astro2020_pathways_2023}, and the Habitable Worlds Observatory (HWO) is currently being designed and optimized for this objective \citep[e.g.,][]{morgan_exo-earth_2023, vaughan_chasing_2023, harada_setting_2024, stark_optimized_2024, stark_paths_2024, mennesson_current_2024, tuchow_hpic_2024}. 
While direct imaging can characterize exoplanet atmospheres, the observation does not directly provide information on the world's mass, radius, orbital phase, inclination, or eccentricity \citep{nayak_atmospheric_2017, guimond_direct_2018, ferrer-chavez_biases_2021}. Radius is degenerate with albedo in reflected light, and radius constraints from transits are unlikely \citep{cahoy_exoplanet_2010, nayak_atmospheric_2017, lustig-yaeger_detecting_2018, gilbert-janizek_retrieved_2024, tuchow_bioverse_2025}, adding uncertainty to future direct imaging observations that use reflected light spectra. 
Radial velocity observations could provide minimum mass and, in combination with assumptions about planet composition, rough estimates of planetary radius. However, the quality of these estimates will depend on system inclination, which is degenerate with eccentricity in direct imaging \citep{ferrer-chavez_biases_2021, damiano_effects_2025}. It should also be noted that radial velocity observations of Earth-Sun combinations are currently not possible, although upcoming extremely large telescopes may be able to do this (see e.g., \cite{gupta_fishing_2024}). These unknowns make it extremely challenging to infer habitability from reflected light spectra without additional information on the system.

\begin{figure}[ht!]
\includegraphics[width=0.97\textwidth]{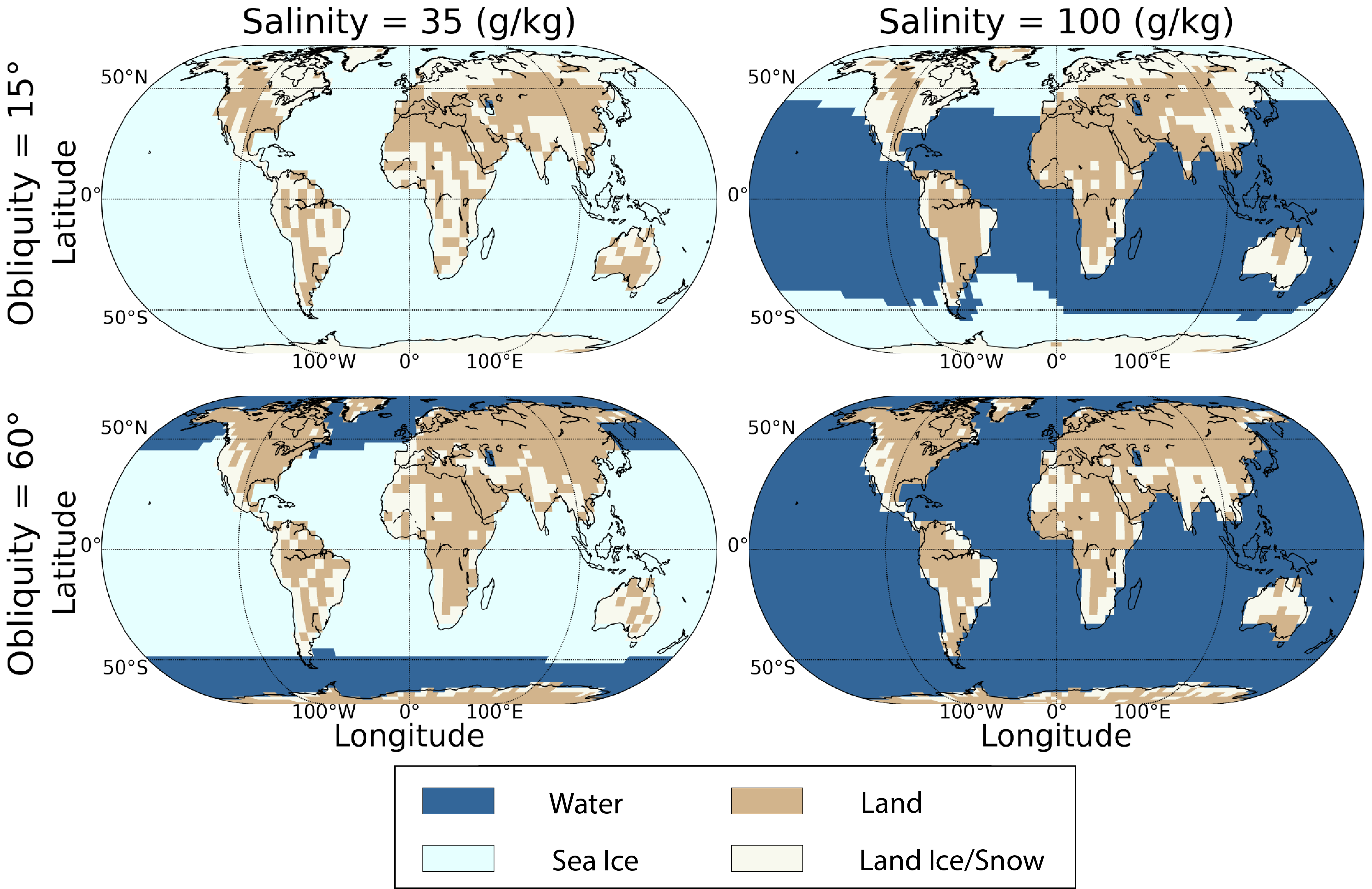}
\caption{Global average climate maps of G-star exoplanets at {low} instellation (0.814 $S/S_{o}$) {simulated} with ocean salinities of 35 {and} 100 g/kg, from left to right, and planetary obliquity {of} 15 {and} 60$^{\circ}$, from top to bottom. We map the decadal mean ocean and continent landmass for each simulation showing gridcells where sea ice or land ice and snow are present. Figure adapted from data in \cite{batra_synergistic_2026}.
\label{fig:ArchIcePlot}}
\end{figure}

The measured apparent albedo of an exoplanet is contingent on the world's atmospheric, cloud, and surface conditions \citep{donohoe_atmospheric_2011, roccetti_planet_2025}. The effect that unresolvable planetary parameters have on climate may thus have indirect but measurable impacts on the resulting spectrum that could be detected with future HWO observations \citep{kofman_pale_2024, guzewich_impact_2020}. 
We explore this possibility by performing simulated observations of four planets with same instellation and atmospheric composition but different climate states (Figure \ref{fig:ArchIcePlot}), including globally glaciated, ice belt (equatorial ice), ice cap (ice at the poles), and ice-free climate states \citep{batra_synergistic_2026}.

\section{Methods} \label{sec:methods}
\subsection{General Circulation Model Experiments}

We simulate observations of four end-member climate states modeled by \cite{batra_synergistic_2026}. These simulations used the ROCKE-3D, a versatile exoplanet General Circulation Model (GCM) developed by NASA GISS \citep{way_resolving_2017} and we extend experiments a few decades using a higher resolution SOCRATES radiation scheme for spectral analysis \citep{edwards_studies_1996, edwards_efficient_1996}. All four simulations have a 1 bar N$_2$-dominated atmosphere resembling {present-day} Earth (78.1\% N$_2$ and 20.9\% O$_2$) with 285.2 ppm atmospheric CO$_{2}$ (pre-industrial) and receive {low} Archean{-like} instellation (1,108 W/m$^2$ or 0.814 S$_\earth$). The {distinct climate states and ice cover in each GCM experiment arise} from differences in ocean salinity and planetary obliquity {imposed in each simulation} (Figure \ref{fig:ArchIcePlot}){, rather than prescribing climate state \citep{batra_synergistic_2026}}. {The salinity and obliquity used in these simulations create a two by two `factorial design' experiment setup, where each salinity is run at both obliquities and each obliquity is explored at both ocean salinities.} We define the {resulting} climate state of the experiment with 35 g/kg salinity and 15$^{\circ}$ planetary obliquity as globally glaciated, the experiment with 35 g/kg salinity and 60$^{\circ}$ planetary obliquity as an ice belt state, the experiment with 100 g/kg salinity and 15$^{\circ}$ planetary obliquity as an ice cap state, and the experiment with 100 g/kg salinity and 60$^{\circ}$ planetary obliquity as ice-free. {The experiments are not paleo-Earth analogs, rather they isolate the impact of climate state at a lower, Archean-like, instellation with a fixed atmospheric composition, which allows for greater variability in ice cover and albedo compared to warmer climates with higher flux.} Additional details regarding ROCKE-3D and the previously published {experiments exploring how salinity and obliquity impact climate state that} we consider are provided by \cite{way_resolving_2017} and \cite{batra_synergistic_2026}, respectively.

\subsection{Planetary Spectrum Generator}
We use the Planetary Spectrum Generator (PSG) \citep{villanueva_planetary_2018} spectroscopic suite to simulate synthetic reflectance spectra of each of the GCM simulation we adapt. PSG is an online radiative transfer tool capable of generating synthetic spectra of planetary atmospheres over a broad range of wavelengths and observatory configurations.
We use the Global Emission Spectra (GlobES) application, the 3D framework of PSG, allowing for the input of our atmospheric and surface GCM data into the tool to generate reflectance spectra of the modeled world \citep{kofman_pale_2024, fauchez_global_2025}. 
Further details on PSG, the GlobES app, and the input of GCM and configuration files to generate spectra can be found in the PSG Handbook \citep{villanueva_fundamentals_2022} as well as in \cite{villanueva_planetary_2018} and \cite{fauchez_global_2025}. 
The GlobES application has previously been used to synthesize exoplanetary spectra from a variety of GCMs including ROCKE-3D \citep[e.g.,][]{fauchez_trappist-1_2022, kofman_pale_2024, metz_detectability_2024, hammond_climates_2025, kelkar_earth_2025}. 

We simulate all spectra with the recommended HWO telescope configuration (3.73 L/D inner working angle versus throughput) from the most recent astronomy and astrophysics decadal survey \citep{committee_for_a_decadal_survey_on_astronomy_and_astrophysics_2020_astro2020_pathways_2023}. No rotation is considered in observations, so the spectra of each world are solely from the observer's perspective and held consistent at each obliquity. Previous work suggests that for rapidly rotating planets where planetary variation is overwhelmingly latitudinal (like Earth) there are only minor changes in apparent albedo spectra at different points in the observation \citep{kofman_pale_2024}.
The behavior of clouds on exoplanets is a major source of uncertainty in modeling and future observations \citep{schwieterman_exoplanet_2018, helling_exoplanet_2019} and clouds have nonlinearity in reflectivity and spatial abundance/inhomogeneity not fully captured at the resolution of our GCM experiments \citep{roccetti_planet_2025}. To account for this, we generate annual mean spectra of each GCM simulation {in PSG} both without and with water and ice clouds. {We remove the clouds from the PSG-computed spectra as an artificial contrast experiment to isolate the contribution of gaseous absorption and albedo to the spectra. These experiments are not representative of worlds with a cloud-free climate, as planets without clouds would have an altered climate evolution. While PSG includes advanced cloud radiative-transfer capabilities, there is no straightforward way to consider clouds in any radiative-transfer model due to uncertainties regarding clouds on other worlds \citep{villanueva_fundamentals_2022, kofman_pale_2024}. The diagnostic experiments where clouds are removed therefore provide a lower limit on planetary albedo, since clouds generally increase planetary reflectivity.}
We also consider an annual mean case with lower pO$_{2}$ (10\% PAL O$_{2}$), an upper estimate of `Proterozoic-like' oxygen \citep{olson_earth_2018}, with an otherwise unchanged atmosphere {(not an analog) to investigate climate state impacts on distinct oxygen features}. 
We then explore spectra of the GCM simulations with four monthly mean outputs (at present-day O$_{2}$), encompassing the winter solstice, vernal equinox, summer solstice, and autumnal equinox respectively to compare to the mean spectrum of each climate state as well as investigate the seasonal variability of the spectra caused by planetary obliquity.
{We generate reflectance spectra as apparent albedos [I/F], defined as the flux ratio of observed planetary brightness compared to an idealized fully reflecting flat (planet-sized) disk \citep{villanueva_fundamentals_2022}, plotted} over a wavelength range of 250-1700 nm for each simulation test case. This is a wider wavelength range than that of the recommended HWO telescope design, but research to determine optimal observation ranges is still under discussion and expanding our synthetic observations ensures that no features are neglected \citep{arney_search_2025, krissansen-totton_wavelength_2025}. 

The strength the signal is expressed as its `equivalent width' which is determined by linearly integrating between the start and the end of the spectral feature. The units of equivalent width are nm, and they represent the width a feature would have if it extended from the continuum level to zero. We calculate the equivalent width of the integrated spectral bands of O$_{2}$ (750-778 nm {bandpass}) and H$_{2}$O (787-857 nm {bandpass}) absorption. 
To calculate how long it would take to detect O$_{2}$ {(over the 750-778 nm bandpass)}, we use the signal-to-noise ratio (S/N) of that feature within the spectra using prescribed telescope design and photon noise (shot noise). {S/N uses the resolving power of the prescribed instrument, which determines the bin width. Each bin collects a photon count over the observation time and the noise is the square root of the photon count, thus doubling observation time increases the S/N by 1.44.}
We calculate the time to reach a S/N $\geq$ 10 at a distance of 10 pc using a python-based noise estimation code adapted from the PSG noise model \cite{kofman_pale_2024}, found in the NASA PSG github (\url{https://github.com/nasapsg/globes/blob/main/pyglobes/noise_parameterization.py}){, and further details on S/N of PSG observations can be found in the PSG Handbook \citep{villanueva_fundamentals_2022}}. {Our exposure time calculations assume the same wavelength band and sampling assumptions as the equivalent width measurement.}
We calculate exposure times with a 6 m mirror, as recommended by the decadal survey \citep{committee_for_a_decadal_survey_on_astronomy_and_astrophysics_2020_astro2020_pathways_2023}, as well as an 8 m mirror. All calculations are completed both without and with clouds.


\section{Results} \label{sec:results}

\begin{figure}[ht!]
\includegraphics[width=0.97\textwidth]{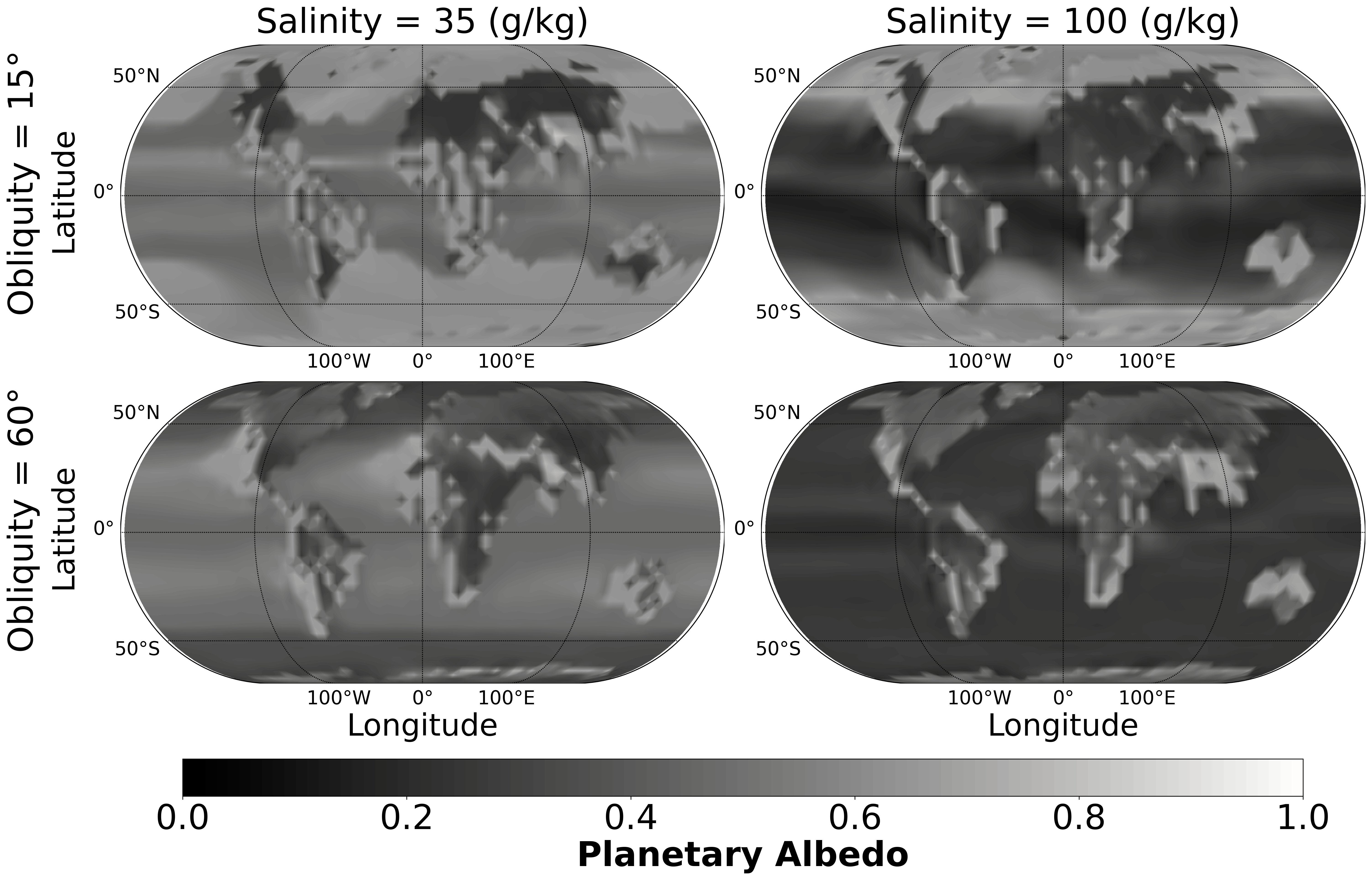}
\caption{Global wavelength-integrated {decadally averaged} planetary albedo (ratio of reflected to incident shortwave radiation at the top of the atmosphere) maps from 0 to 1 for G-star exoplanets at {low} instellation (0.814 $S/S_{o}$) {simulated} with ocean salinities of 35 {and} 100 g/kg, from left to right, and planetary obliquity {of} 15 {and} 60$^{\circ}$, from top to bottom.
\label{fig:ArchAlbedoPlot}}
\end{figure}

\begin{figure}[ht!]
\includegraphics[width=1.0\textwidth]{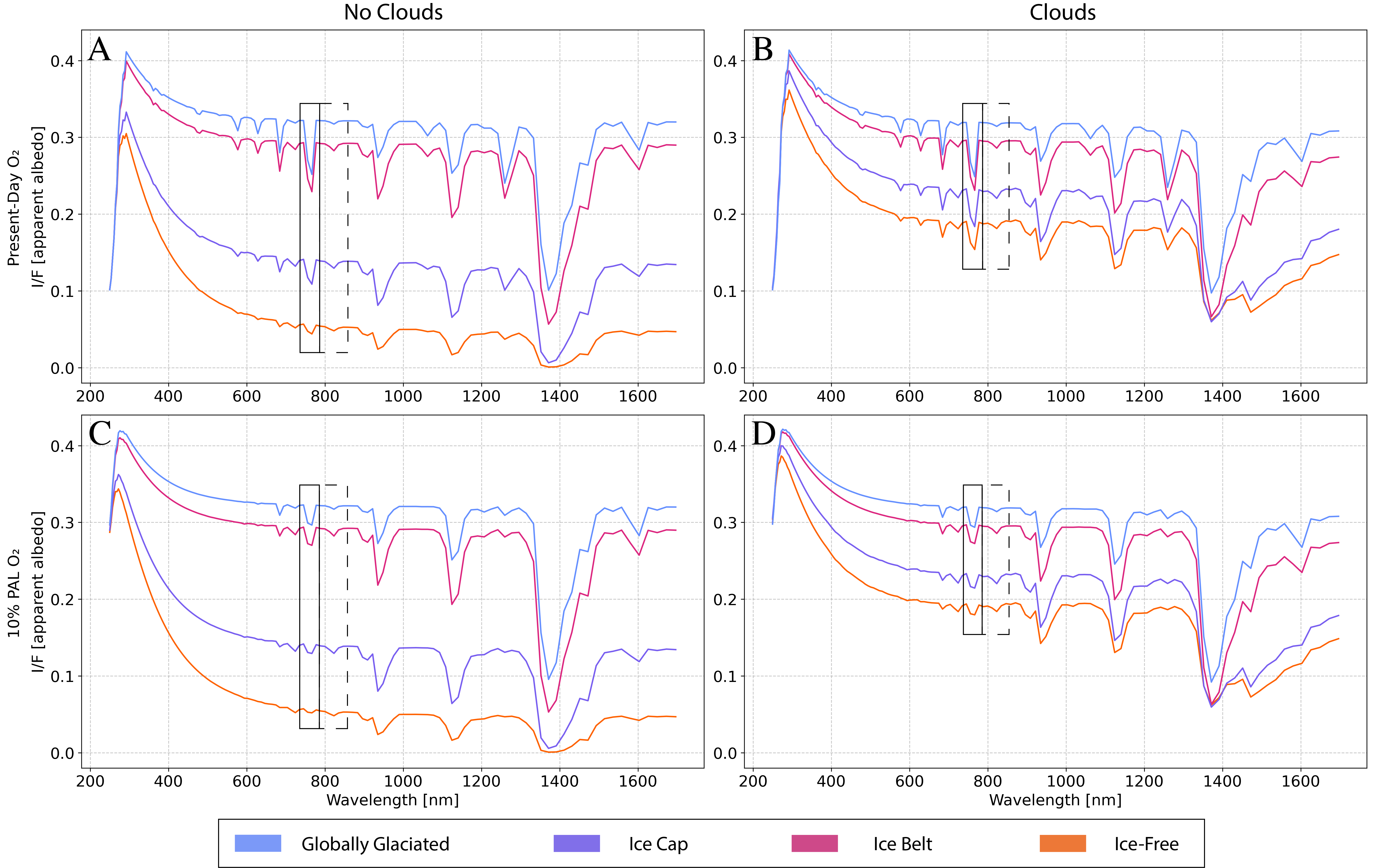}
\caption{Simulated spectra of annual mean GCM experiments at each climate state (A) without clouds and (B) with water and ice clouds.
Below, we map the annual mean spectra of each experiment with low, 10\% PAL, O$_{2}$ (C) without clouds and (D) with water and ice clouds. The atmosphere is otherwise unchanged from A and B. {Solid black rectangle highlights the O$_{2}$ (750-778 nm) feature of each spectra, while dashed black rectangle highlights the H$_{2}$O (787-857 nm) feature of each spectra. The assumed spectral resolving power used is R70.}
\label{fig:HWOCloudProt}}
\end{figure}

\begin{figure}[ht!]
\centering
\includegraphics[width=0.99\textwidth]{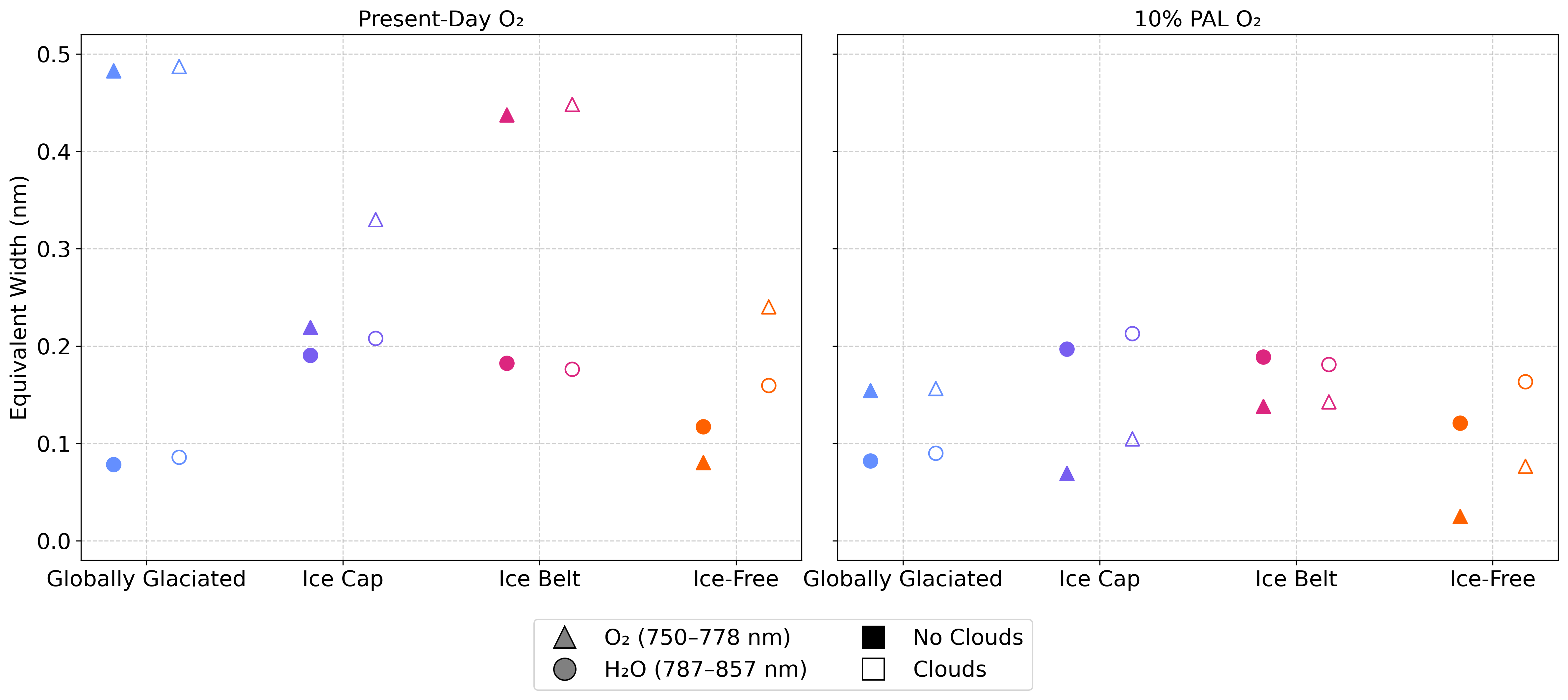}
\caption{Equivalent widths of O$_{2}$ (750-778 nm) and H$_{2}$O (787-857 nm) spectral features of our annual mean GCM experiments (with present-day atmospheric O$_{2}$) at each climate state without clouds and with water and ice clouds (left). The equivalent widths of the same features for worlds {with} 10\% PAL O$_{2}$ to resemble a low oxygen world (right).
\label{fig:HWO_EW}}
\end{figure}

\begin{figure}[ht!]
\includegraphics[width=1.0\textwidth]{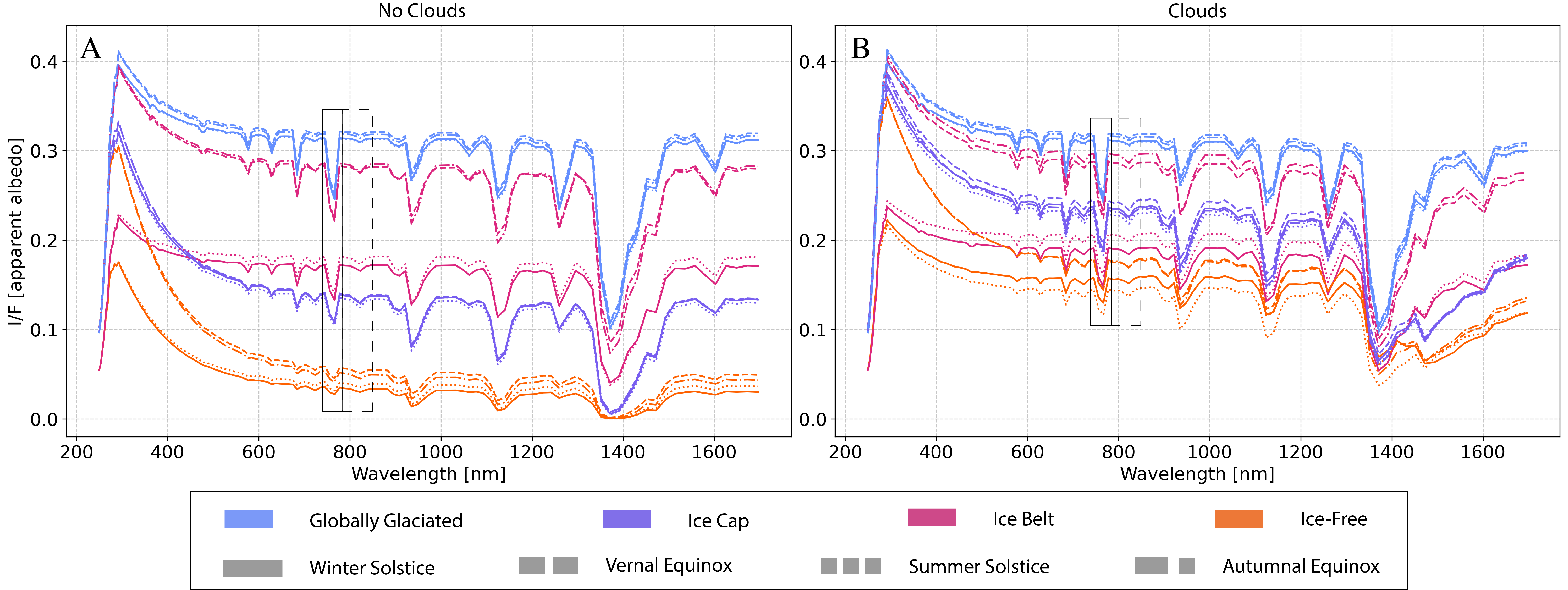}
\caption{Simulated seasonal monthly mean spectra encompassing the equinoxes and solstices of GCM experiments at each climate state (A) without clouds and (B) with water and ice clouds. {Solid black rectangle highlights the O$_{2}$ (750-778 nm) feature of each spectra, while dashed black rectangle highlights the H$_{2}$O (787-857 nm) feature of each spectra. The assumed spectral resolving power used is R70.}
\label{fig:HWOCloudNormSeasons}}
\end{figure}

\begin{figure}[ht!]
\includegraphics[width=1.0\textwidth]{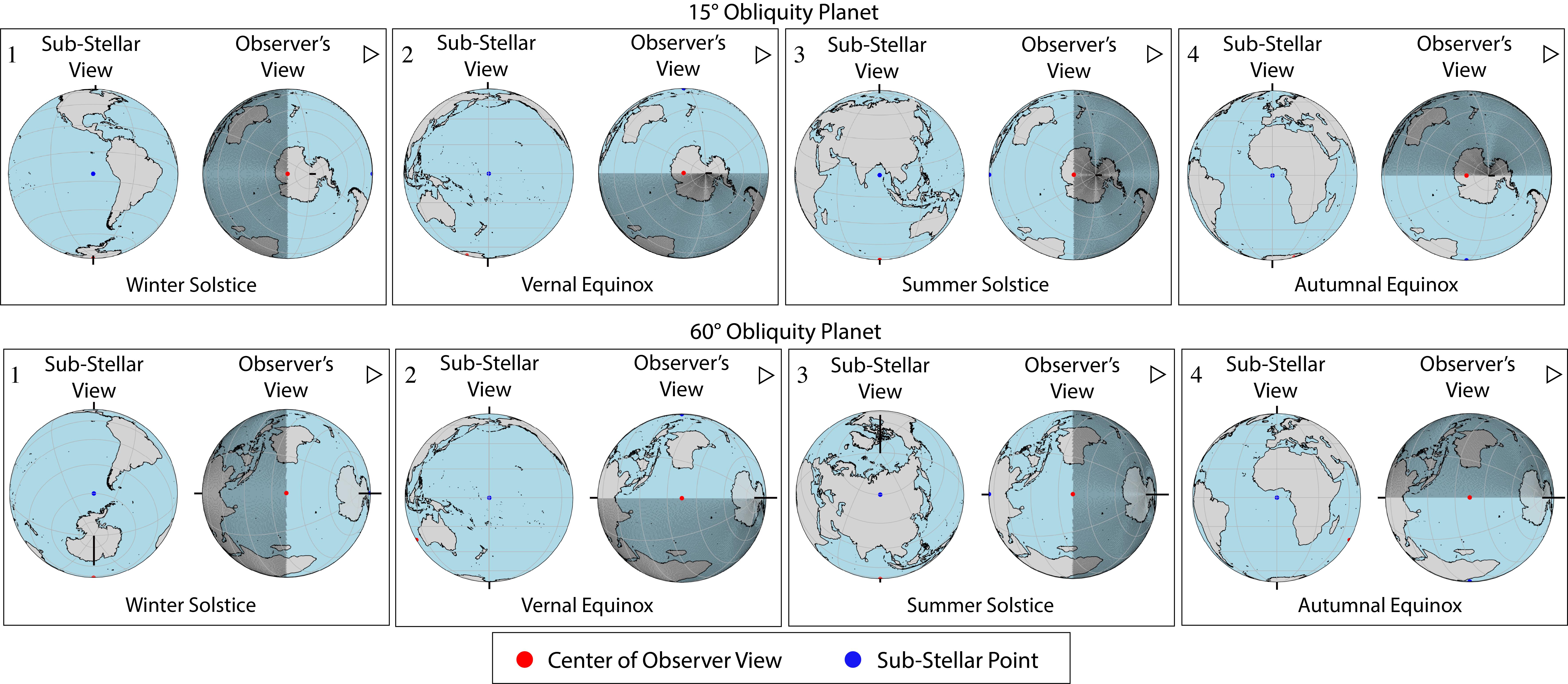}
\caption{Cartoon of the at quadrature observations of a planet from the perspective of the sub-stellar point and of an observer at each equinox and solstice for a 15$^{\circ}$ and 60$^{\circ}$ obliquity world. The observer's view is held fixed for consistency (rather than a fixed sub-stellar point) in each panel and the obliquities shown are consistent with the climate state experiments we explore.
\label{fig:cartoon}}
\end{figure}

\begin{figure}[ht!]
\centering
\includegraphics[width=0.95\textwidth]{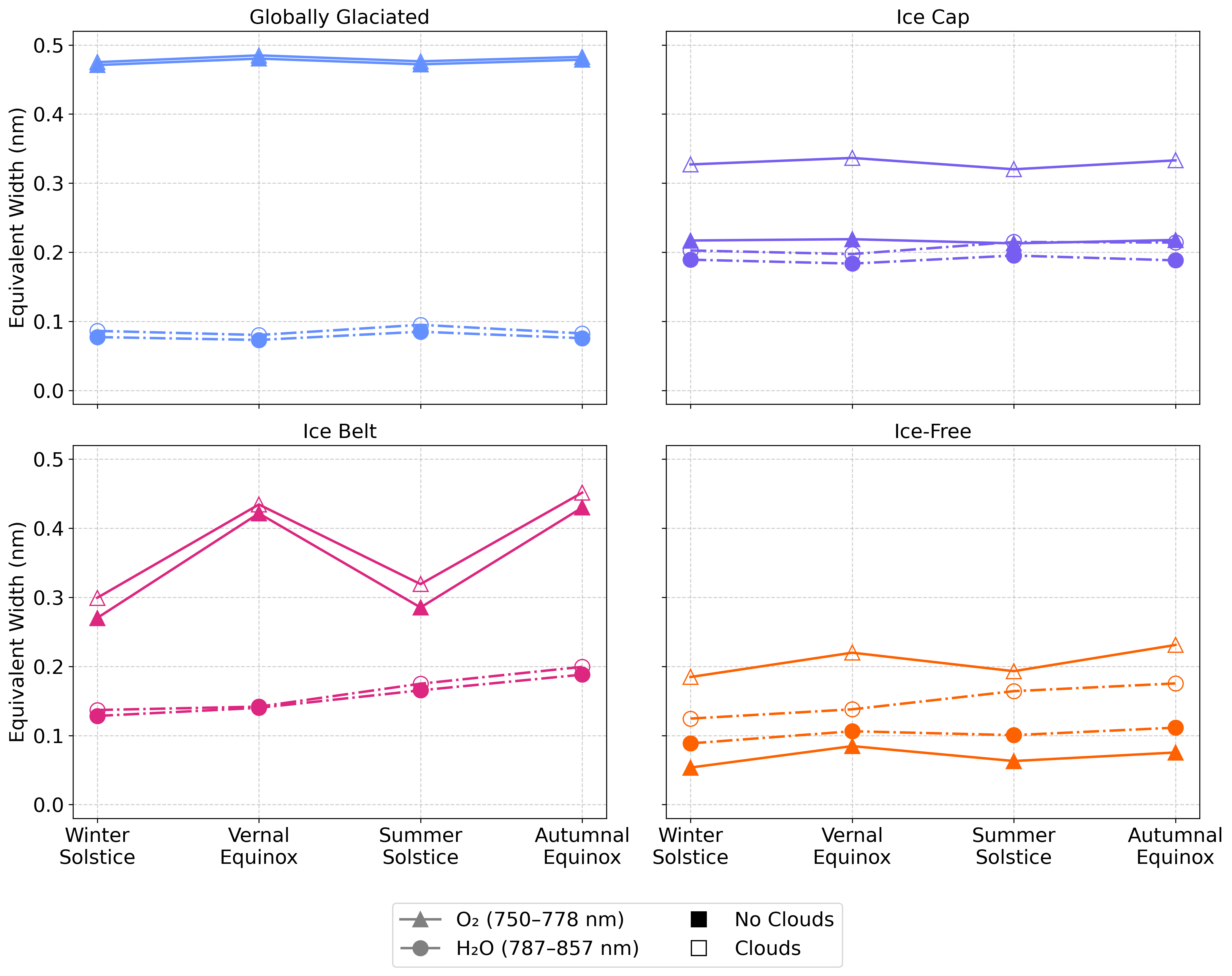}
\caption{Equivalent widths of O$_{2}$ (750-778 nm) and H$_{2}$O (787-857 nm) spectral features of the monthly mean spectra at equinoxes and solstices for each climate state, without clouds and with water and ice clouds.
\label{fig:HWO_EW_Seasons}}
\end{figure}

The climate states we explore are shown in Figure \ref{fig:ArchIcePlot} where we map each world's ocean, sea ice extent, continents, and land snow and ice cover. The relative albedo (reflectance) of each climate state is mapped in Figure \ref{fig:ArchAlbedoPlot}. 

\subsection{Annual Mean Spectra} \label{Subsec:Annual}
We find that the climate states of our experiments impact planetary reflectivity through changes to surface snow/ice extent and distribution, and thus affect the apparent albedo reflectance spectra of those worlds. We generate annual mean reflected light spectra for each climate state without and with clouds, as shown in Figure \ref{fig:HWOCloudProt}, for present-day Earth O$_{2}$. 
Without clouds, we find that the variations in surface reflectivity on the four worlds (Figure \ref{fig:ArchAlbedoPlot}) control the overall brightness of their spectra (Figure \ref{fig:HWOCloudProt}A). Higher surface albedo results in the maximum at the Rayleigh peak (near $\sim$300 nm) being significantly higher ($\sim$0.4 apparent albedo) in the glaciated planets versus those that are water dominated. The ice belt world has the second brightest average spectrum, while the ice cap planet apparent albedo is third highest. The ice-free planet has a Rayleigh peak of $\sim$0.3 and its apparent albedo is considerably lower than the other three climate states across the entire spectrum. 

Clouds reflect light and thus increase the brightness and apparent albedo of all climate scenarios with clouds compared to those without clouds (Figure \ref{fig:HWOCloudProt}B). However, the increase in brightness with clouds present is not consistent across all four experiments. The lower albedo worlds (ice-free and ice cap) have a larger increase in apparent albedo when clouds are present compared to their spectra without clouds than the high albedo worlds. Even with the addition of clouds, the reflectivity and brightness of the spectra of planets with more surface ice is greater than those with less ice. 

Next, we simulate the apparent albedo at each climate state with 10\% present atmospheric level (PAL) O$_2$ without clouds (Figure \ref{fig:HWOCloudProt}C) and with clouds (D) to investigate low oxygen worlds (the atmosphere is otherwise unchanged). The same trends of climate state impacting the brightness of their apparent albedo hold true for the worlds with low O$_{2}$ (10\% PAL) without and with clouds (Figure \ref{fig:HWOCloudProt}C and D), as they have the same ice cover and clouds as the higher O$_{2}$ experiments. 

We find that the strength of atmospheric features within the spectra are dependent on the planet's climate state reflectivity (Figure \ref{fig:HWO_EW}), calculated by removing the feature in annual mean spectra through a linear interpolation {(relative to its local continuum)} to determine equivalent width. We explore the O$_{2}$ (750-778 nm) and H$_{2}$O (787-857 nm) {bandpass} features, {gases} relevant in Earth's atmosphere that are potentially indicators of exoplanet habitability and are useful for the interpretation of exoplanet biosignatures \citep{schwieterman_exoplanet_2018, schwieterman_overview_2024, kofman_pale_2024}. 
The strength of features{,} determined by the reflectivity of the observed world{,} impact the total amount of photons received (between different climate states and temporally on a planet){. The equivalent width, as well as the total strength of the signal, will therefore impact the time it takes to detect and confidently identify a specific atmospheric feature in observations, assuming identical wavelength band definitions and sampling assumptions \citep{young_inferring_2024, kofman_pale_2024}.}
The O$_{2}$ and H$_{2}$O features typically have greater equivalent widths on planets with higher albedos, i.e. the ones with greater ice cover. Features are stronger when there is more light reflected from the planets surface, like for the globally glaciated and ice belt worlds (Figure \ref{fig:HWO_EW}A). For present-day Earth pO$_2$, the equivalent width of the O$_{2}$ feature is $\sim$0.48 nm and the H$_{2}$O feature is $\sim$0.08 nm on the globally glaciated world without clouds. The ice cap world (without clouds) has a lower O$_{2}$ equivalent width ($\sim$0.22 nm) but a larger H$_{2}$O feature (0.19 nm), as water vapor is higher in simulations with surface liquid water. The ice belt world has a greater O$_{2}$ feature equivalent width compared to the ice cap world ($\sim$0.43 nm), while the H$_{2}$O feature is comparable ($\sim$0.18 nm). On the ice-free world without clouds, the equivalent width of the O$_{2}$ feature ($\sim$0.08 nm) is smaller than the other three climate states and the H$_{2}$O feature ($\sim$0.12 nm) is weaker than the ice belt and ice cap worlds (though the H$_{2}$O feature is greater than that of the globally glaciated world). 
With water and ice clouds, the equivalent widths of most atmospheric features are slightly higher (Figure \ref{fig:HWO_EW}A). Clouds do not have a major an impact on the apparent albedo and feature equivalent widths for planets with high albedos already (globally glaciated and ice belt worlds). However, clouds do have a significant impact on planets with lower albedos (the ice cap and ice-free worlds), markedly increasing their annual mean O$_{2}$ and H$_{2}$O feature strengths. For the ice-free world, O$_{2}$ has a $3\times$ greater feature strength and H$_{2}$O is $1.3\times$ greater with clouds compared to without.

We also find the feature strength of low O$_{2}$ planets (10\% PAL), calculating the equivalent widths of the features (without and with clouds) for each climate state in Figure \ref{fig:HWO_EW}B. All other atmospheric abundances, including that of water vapor, are unchanged from the present-day pO$_2$ calculations. The O$_{2}$ equivalent widths are significantly reduced at 10\% PAL O$_{2}$ compared to present-day oxygen for all four climate states. The trends in O$_{2}$ feature strength across low O$_{2}$ scenarios are consistent with the results at present-day O$_{2}$, where the globally glaciated world has the greatest O$_{2}$ strength ($\sim$0.15 nm) and the ice belt world has the next strongest feature ($\sim$0.14 nm). The two ice cap and ice-free worlds have lower O$_{2}$ features (below 0.1 nm), but experience a larger increase in equivalent width when clouds are included than high albedo worlds (Figure \ref{fig:HWO_EW}B). 

\subsection{Monthly Mean Spectra} \label{Subsec:Seasonal}
We find that high obliquity worlds have a variation in spectral brightness across different seasons, especially significant in the ice belt scenario. We record four monthly mean spectra (with present-day O$_{2}$) of at quadrature observations of our four climate states (Figure \ref{fig:HWOCloudNormSeasons}) without (A) and with (B) clouds. We show a cartoon example of the sub-stellar and observer's perspective of at quadrature observations at each solstice and equinox for 15$^{\circ}$ and 60$^{\circ}$ obliquity worlds in Figure \ref{fig:cartoon} for reference, which represent the orbital configurations of the generated spectra in Figure \ref{fig:HWOCloudNormSeasons}.
The trends observed in the annual mean spectra, where the globally glaciated world has the highest apparent albedo across its spectrum while the ice-free world has the lowest, still hold true at each season. However, we also find temporal variations in the monthly mean spectra across the seasons for each individual climate state. For the globally glaciated and ice cap worlds (low obliquity experiments) without clouds there is a minor variation in the Rayleigh peak ($\sim$ 300 nm) of the monthly mean spectra between the solstices and equinoxes, where the (northern hemisphere) winter and summer solstice have lower peak apparent albedos. At greater wavelengths, the apparent albedo at each season approach similar values for each respective world (Figure \ref{fig:HWOCloudNormSeasons}A).
For the ice-free world (at high obliquity), the winter and summer solstice monthly mean spectra have a Rayleigh peak that is significantly lower than that of the peaks at both equinoxes. The disparity in apparent albedo is smaller at greater wavelengths, but there is still a difference between the equinox and solstice spectra. In the ice belt experiment (at high obliquity), not only do we observe a major difference between the Rayleigh peaks of solstice and equinox experiments, there is a significant variation in strength across the entire spectra where the solstices are always lower than the equinoxes. 

We find that the monthly mean spectra at every climate state with clouds (Figure \ref{fig:HWOCloudNormSeasons}B) follows similar trends to the spectra to without clouds (Figure \ref{fig:HWOCloudNormSeasons}A), except all spectra increase in apparent albedo. We observe the same seasonal disparity in the apparent albedo of the Rayleigh peak between the solstice and equinox monthly mean spectra for high obliquity experiments (ice-free and ice belt worlds), while low obliquity worlds have only a slight difference (Figure \ref{fig:HWOCloudNormSeasons}B). The ice belt world has significant seasonal contrast across its entire spectra in particular, where the equinoxes are brighter than the solstices.
All four ice cap monthly mean spectra are brighter than the ice belt solstice apparent albedos due to clouds raising reflectivity on the ice cap world. This is not the case for the experiments without clouds, where the ice cap reflectivity is lower than the ice belt world apparent albedo at every season.

We explore the temporal variability of the O$_{2}$ and H$_{2}$O spectral feature equivalent widths (nm) for our monthly averaged solstice and equinox experiments at each climate state (Figure \ref{fig:HWO_EW_Seasons}) without clouds and with water and ice clouds. The feature equivalent width trends discussed in the annual mean results (Section \ref{Subsec:Annual}) hold true for our monthly mean data when comparing between different climate state experiments for a given season. The worlds with greater ice cover have the strongest O$_{2}$ feature while clouds impact the equivalent widths of the low albedo worlds more significantly. 
However, we also find temporal variation in the feature strength between seasons in our monthly mean climate state experiments (Figure \ref{fig:HWO_EW_Seasons}). The globally glaciated and ice cap world O$_{2}$ and H$_{2}$O features change only slightly in strength. The change in equivalent width across different seasons is small in the low obliquity worlds ($\sim$0.01 nm both without and with clouds).
We observe a greater variability in feature size temporally in our high obliquity experiments (Figure \ref{fig:HWO_EW_Seasons}). The ice belt world without clouds has a significant change in O$_{2}$ feature strength, with O$_{2}$ equivalent width near $\sim$0.28 nm at the solstices and $\sim$0.43 nm at the equinoxes. The O$_{2}$ feature on the ice-free world without clouds displays a similar trend of greater strength during the equinoxes while smaller at the solstices, though the variation is lower in magnitude than the ice belt world. The H$_{2}$O feature increases temporally over the year from winter solstice to autumnal equinox on the high obliquity worlds. 

Clouds increase the brightness and feature equivalent widths for all monthly mean scenarios, but most significantly for the ice cap and ice-free worlds with less ice. For example, the O$_{2}$ feature in the ice-free world with clouds is $>$3$\times$ greater and the H$_{2}$O strength is $>$1.3$\times$ at every seasonal monthly mean spectra than without clouds. 
We find that clouds also increase the change in strength between solstice and equinox for our ice-free (high obliquity) scenario, leading to a larger temporal disparity of the equivalent width in spectroscopic observations with clouds compared to without (Figure \ref{fig:HWO_EW_Seasons}). Due to this, not only are the ice-free world features stronger with clouds, but they vary temporally to a greater extent as well.

\section{Discussion} \label{sec:discussion}
We discuss the impact climate state has on the resulting annual mean spectra in Subsection \ref{subsec:state} and the effect clouds in Subsection \ref{subsec:clouds}. We then examine the temporal variability of reflectance spectra in Subsection \ref{subsec:season} and consider the implications it has for searching for biosignatures in Subsection \ref{subsec:Biosigs}. We discuss how planetary climate state impacts observation times in observation of reflectance spectra in Subsection \ref{subsec:error}.

\subsection{Climate State Influence on Observations} \label{subsec:state}
Earth-like exoplanets in the habitable zone can be stable in a large diversity of habitable climate states, ranging between fully ice-free to globally glaciated. Our four experiments are an example of how exoplanets could end up in distinctly different climates, even at {low (Archean-like)} instellation where we might anticipate glaciation (Figure \ref{fig:ArchIcePlot}). Other climates states are possible at different instellations (such as present-day instellation like in \cite{batra_synergistic_2026}) or with entirely different planetary parameters impacting climate.
We find that there is a distinct difference in the spectrum of each climate state in our annual mean scenarios, despite each world having identical atmospheric compositions (Figure \ref{fig:HWOCloudProt}). The apparent albedo of icy worlds (like the globally glaciated and ice belt planets) are higher than those with less ice across the entire reflectance spectrum. The ice-free world has the lowest apparent albedo spectra in our experiments, even with clouds.
Characterizing climate and habitability in reflected light is a major challenge and just observing the mean spectra would likely be insufficient to determine a world's climate state given the radius-albedo degeneracy.
However, we suggest that climate states such as those exemplified by our study may be discernible in future spectroscopic observations if additional information on the planet is known, as our experiments show that there is a distinguishable difference in annual mean apparent albedo between the four states.
Understanding how spectra are affected by planetary climate may reduce uncertainty in the interpretation of future direct imaging measurements of exoplanets using reflected light, potentially allowing for the classification of climate state and improving the interpretation of habitability.

Icy {worlds} also have stronger spectroscopic features than worlds with the limited or no ice cover due to photon reflection, even when the gas abundances in their atmospheres are identical. The globally glaciated world has the greatest O$_{2}$ feature strength, and the O$_{2}$ and H$_{2}$O bands are larger on the ice cap and ice belt worlds than in the ice-free spectra (Figure \ref{fig:HWO_EW}{A-B}). Despite the ice-free world having the greatest ocean fractional habitability of the climate states we consider \citep{batra_synergistic_2026}, the smaller relative feature strengths of O$_{2}$ and H$_{2}$O may make it more difficult to search for biosignatures {(assuming observations use the same sampling assumptions and wavelength band definition)} \citep{schwieterman_overview_2024}. On the other hand, although we fixed pO$_2$ between climate states, global glaciation may limit sea-air gas exchange and the atmospheric O$_2$ accumulation \citep{else_wintertime_2011, else_annual_2013, loose_parameter_2014}. For this reason, high obliquity ice belt planets may be most favorable for biosignature accumulation and detection. We expect that icy, high albedo exoplanets will likely have greater feature strengths than equivalent low albedo worlds for other biosignature {gases} that we did not examine, such as O$_{3}$ and CH$_{4}$.

\subsection{Impact of Clouds} \label{subsec:clouds}

The effect clouds have on the detectability of atmospheric features is of major concern for current and future spectroscopic exoplanet observations \citep{rugheimer_spectral_2013, schwieterman_exoplanet_2018, catling_exoplanet_2018, helling_exoplanet_2019}. Clouds create uncertainty when their height/location, opacity, and abundance are unknown \citep{yang_simulations_2019, kawashima_theoretical_2019, checlair_probing_2021, roccetti_planet_2025}, which may impact the strength of atmospheric signals. High clouds could obscure {gases} below them reducing detectability, while absorption features of {gases} above the clouds can be amplified by cloud reflectivity making reflected light direct imaging observations more sensitive to their features \citep{wang_baseline_2018, kofman_pale_2024, yang_clouds_2025}. Lower spatial resolutions resolved in exoplanet GCMs and model assumptions on cloud variability can create further uncertainties in experiments \citep{roccetti_planet_2025}.
The climate state experiments we explore were generated with a model shown to be able to reproduce major cloud features and incorporate radiative properties of clouds that are not accurately represented in 1D models \citep{fauchez_trappist_2021, sergeev_trappist-1_2022, kofman_pale_2024}. The impact the spatial abundances of clouds have on spectra and atmospheric features in our experiments can be used as an example of what to expect on an exoplanet, though further work is needed to improve the understanding of cloud abundance, behavior, and their effect on recorded spectra. 
We investigate the impact clouds have on spectroscopic observations using our four climate state GCM experiments while also comparing to baseline spectra which do not consider clouds. The ice cap and ice-free {worlds} with the greatest fractional habitability {\citep{batra_synergistic_2026}} have the lowest apparent albedo {(lowest total strength of their signal)}, which hinders the detectability of their atmospheric spectroscopic features. However, clouds notably increase the apparent albedo {of the entire signal} significantly on the low albedo planet, making their spectra brighter and partially mitigating this concern (Figure \ref{fig:HWO_EW}).
O$_{2}$ especially is enhanced with cloudy conditions because it is abundant over the reflective clouds, in agreement with past work \citep{wang_baseline_2018, kawashima_theoretical_2019}, and other {gases} like O$_{3}$ would be amplified by clouds if present as well \citep{kofman_pale_2024, yang_clouds_2025}.

\subsection{Temporal Variability of Spectra} \label{subsec:season}
Not only do worlds with the same atmospheric composition display distinct annual mean reflectance spectra, we find that the monthly mean apparent albedos also differ in temporal variations due to their climate state as well.
Planetary obliquity impacts where the planet is illuminated (Figure \ref{fig:cartoon}), thus the portion of land, ocean, ice, and composition of the atmosphere (location of clouds) that are observed, which determine albedo. This controls brightness of the spectra, creating obliquity driven variations the apparent albedo of observations at different seasons (Figure \ref{fig:HWOCloudNormSeasons}). 
Our ice belt and ice-free worlds therefore have a larger, more significant apparent albedo variation across solstices and equinoxes in spectroscopic observations due to their greater obliquity driven seasonality. The high obliquity of the ice belt world impacts how much of the equatorial ice is observed in at quadrature observations between the solstices and equinoxes, leading to major differences in apparent albedo between observations of the same planet. At the solstices, our view is dominated by the ice-free summer pole; at equinox, the ice belt takes center stage. 
Multiple spectroscopic observations of the temporal contrast in apparent albedo could thus help identify or constrain the climate state and obliquity of an exoplanet {(assuming all observations are at quadrature)}. In contrast with an ice belt planet, an ice cap world with low obliquity would display limited variation across multiple observations in a face-on orbit. 
The equivalent width of atmospheric features like O$_{2}$ are therefore time-varying at solstice vs. equinox spectra (Figure \ref{fig:HWO_EW_Seasons}), for high obliquity worlds. We expect other atmospheric features such as O$_{3}$ and CH$_{4}$ to display similar behavior as well. We also find that the H$_{2}$O feature strength increases from winter solstice to autumnal equinox in Figure \ref{fig:HWOCloudNormSeasons} for high obliquity worlds, likely due hemispheric asymmetry in continent vs. ocean cover across observations compounded with other seasonal changes (see Figure \ref{fig:cartoon}). Multiple spectroscopic observations may thus also help inform if there is hemispheric asymmetry present on the world through mapping apparent albedo \citep{cowan_alien_2009, cowan_rotational_2011, fujii_colors_2010, kawahara_mapping_2011, fujii_mapping_2012, farr_exocartographer_2018}. We also find that clouds amplify the magnitude atmospheric features change in strength temporally compared to without clouds, especially for low albedo but high obliquity worlds (like the ice-free planet), {potentially} positively impacting the detectability of the spectral features.

\begin{figure}[ht!]
\includegraphics[width=0.97\textwidth]{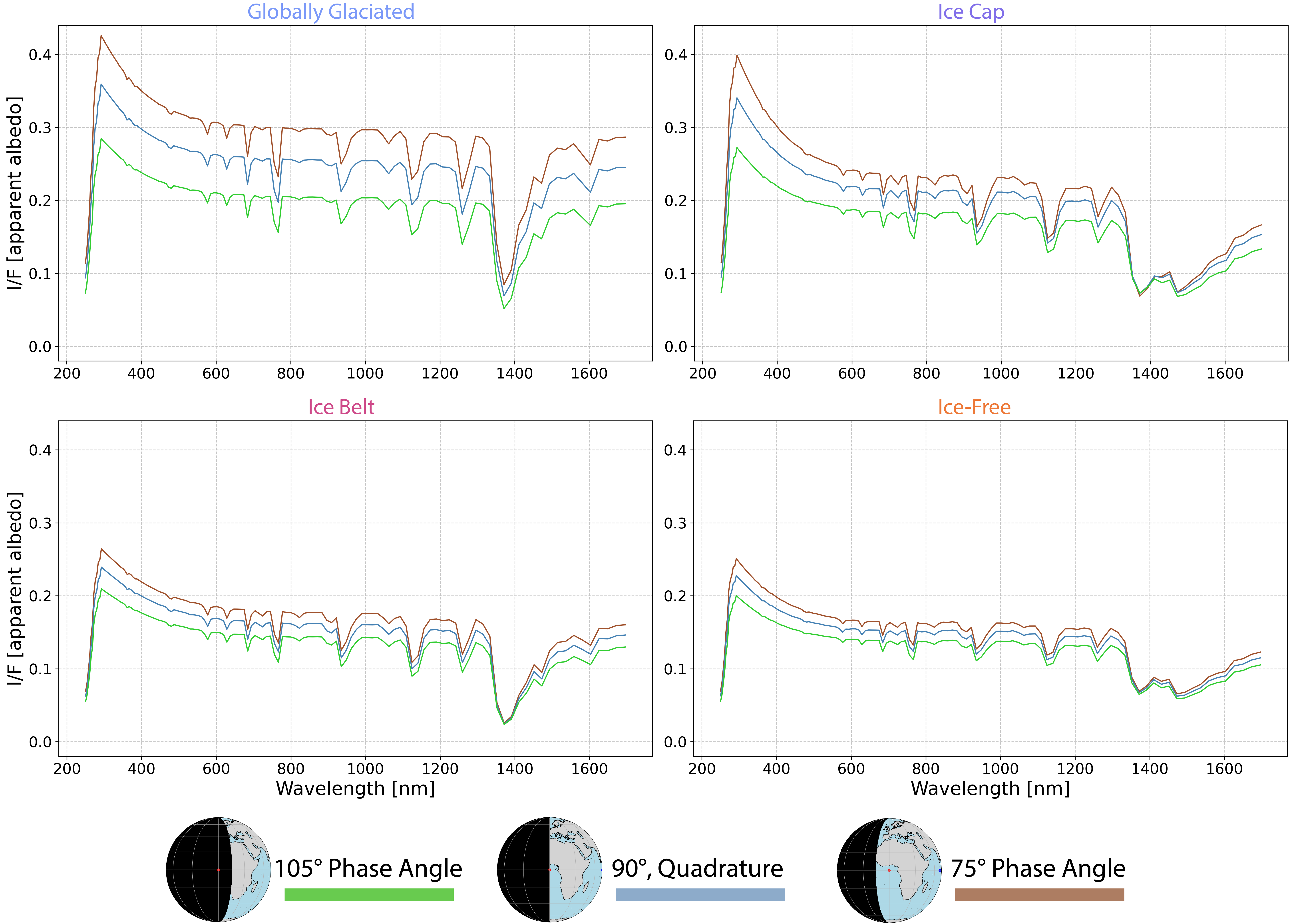}
\caption{{Sensitivity test of the apparent albedo of worlds at quadrature (90$^{\circ}$) as well as gibbous (75$^{\circ}$) and crescent (105$^{\circ}$) phase angles for each climate state world at the winter solstice.}
\label{fig:phase}}
\end{figure}

Moreover, the maximum albedo between repeated observations could help break the variable radius-albedo degeneracy {\citep{salvador_influence_2024, tuchow_bioverse_2025}}, as planetary radius is not temporally variable. A caveat is that this approach requires us to account for changes in phase throughout the planet's orbit, which are modulated by its inclination {and may impact on the resulting spectra observed}. {We generate all our previous spectra with exoplanets at quadrature (assuming a face-on orbit), which maximizes the angular separation between the planet and its host star \citep{vaughan_chasing_2023}. However, it is likely that many exoEarth candidates will not be in face-on orbits and/or will be observed at phases other than quadrature \citep{stark_maximizing_2014, stark_paths_2024, stark_optimized_2024}. Spectroscopic observations that are not fully face-on lead to changes in the proportion of the planet which is illuminated depending on when the observation is taken \citep{robinson_detection_2014, lustig-yaeger_detecting_2018, olson_atmospheric_2018}. Long observation times may also observe the planet while the phase is changing. We generate a phase angle sensitivity test of the apparent albedo of all four climate state worlds (with 200 binning) at the winter solstice of each experiment at quadrature (90$^{\circ}$) as well as a $\pm$15$^{\circ}$ change in phase (relative to quadrature) to simulate observations of gibbous (75$^{\circ}$) and crescent (105$^{\circ}$) phase angle worlds (Figure \ref{fig:phase}). The phase angle of the observed world affects its apparent albedo, with phase-dependent changes in spectral strength corresponding to the planet's reflectivity at the time of observation. The ice-free world has a smaller change in apparent albedo with phase than the globally glaciated world. High obliquity planets like the ice belt world exhibit temporal variability in apparent albedo with phase, depending on what region of the planet is being observed (e.g., poles vs. equatorial latitudes). At the solstices, the change in apparent albedo with phase on the ice belt world is smaller than that of the ice cap world, but at the equinoxes the change is greater, consistent with the seasonal apparent albedo spectra of the ice cap and ice belt worlds in Figure \ref{fig:HWOCloudNormSeasons}. Phase angle can thus lead to significant variability in the entire spectral strength when not at quadrature (Figure \ref{fig:phase}), potentially impacting the detectability of atmospheric features, and must be accounted for when interpreting future reflectance spectra across multiple observations.} 

{Exoplanet orbital eccentricity can have a similar effect varying brightness of spectra temporally across spectroscopic observations as well.} Unfortunately, inclination is degenerate with eccentricity in direct imaging {which may make it difficult to untangle the variation of apparent albedo caused by observational phase from that of seasonality caused by obliquity in observations of worlds that are not observed at quadrature. However, performing concurrent astrometry offers an attractive solution to this problem. Relative astrometry performed concurrently with repeated reflected light observations would provide inclination, which allows for the phase of the world during observations to be inferred \citep{guimond_direct_2018, alei_multi-bandpass_2026}. Knowing the phase of observation, the impact of phase can be measured and separated from any temporal variation in apparent albedo from obliquity, improving albedo constraints. Determining the impact of phase angle on apparent albedo may also help inform planetary climate state, as the sensitivity of apparent albedo to phase is dependent on planetary reflectivity at the time of observation (Figure \ref{fig:phase}).
Isolating the maximum albedo caused by obliquity between repeated observations could help break the variable radius-albedo degeneracy, allowing an estimate of planetary radius. This} would also be an effective way to get a reliable mass estimate of the observed world \citep{damiano_effects_2025, bao_closeby_2025}, further improving our ability to assess planetary habitability with a future direct imaging mission such as HWO. Reducing radius-albedo degeneracy through repeated albedo variation measurements paired with additional information (e.g., from radial velocity or astrometry observations) {to constrain observational phase (if the observed system is not face-on)} could thus help to identify the climate state of Earth-like exoplanets.

\subsection{Seasonal Biosignatures}
\label{subsec:Biosigs}

Seasonality in atmospheric composition of an exoplanet may be a biosignature, as biological activity may vary with seasons leading to temporal changes in the production/consumption of {gases} \citep{aime_direct_2006, meadows_planetary_2008, olson_atmospheric_2018, schwieterman_exoplanet_2018, schwieterman_overview_2024}. Past work has discussed how resulting spectra and the equivalent width of atmospheric {gases} like N$_{2}$O and O$_{3}$ may be time-varying and detectable \citep{olson_atmospheric_2018}. On Earth, the atmosphere has seasonal variations in {gases} such as CO$_{2}$ and O$_{2}$ which are modified by biological activity like respiration and linked to seasonal growth cycles \citep{keeling_global_1996, olson_atmospheric_2018}. Atmospheric seasonality as a biosignature may help mitigate against both false positive and false negative detections. However, how a spectra varies temporally is also dependent on complex factors including viewing geometry and any abiogenic sources that may impact spectroscopic observations over time must be considered \citep{olson_atmospheric_2018, schwieterman_overview_2024}.

We find significant abiogenic seasonality in our monthly mean apparent albedos of high obliquity worlds, {as well as from observations at different phase angles,} which causes atmospheric feature strength to vary temporally even though the atmospheric abundance of {gases} are not changing (Figure \ref{fig:HWOCloudNormSeasons}). This is a distinct difference from a seasonal biosignature signal, where life changes the abundance of specific gases in the atmosphere across different seasons \citep{olson_atmospheric_2018, schwieterman_overview_2024}. 
Increased planetary obliquity amplifies planetary seasonality which may be beneficial for the detection of variable biotic signals, however, it also leads to a greater contrast in the strength of spectra temporally. Features that vary from abiogenic seasonality could be amplified or muted due to life depending on the temporal pattern of biosignatures \citep{olson_atmospheric_2018}. For example, O$_{2}$ production by life on Earth is greatest in the northern summer, while summer solstice O$_{2}$ strength from obliquity is weaker than the equinoxes in our ice belt ($\sim$0.15 nm magnitude without clouds) and ice-free ($\sim$0.03 nm magnitude without clouds) worlds (Figure \ref{fig:HWO_EW_Seasons}), potentially muting the change in feature size we observe. Variations in O$_{2}$ by life at present-day O$_{2}$ abundances may be difficult to detect. However{,} biological seasonality of {gases} at lower atmospheric abundances like Proterozoic O$_{2}$ or other features not explored (e.g., O$_{3}$, CO$_{2}$, CH$_{4}$, N$_{2}$O) may be feasible to detect \citep{olson_atmospheric_2018}, yet would also experience temporal variability from obliquity in reflected light observations, like their behavior in thermal emission spectra \citep{mettler_earth_2023}.
It is thus important for future spectroscopic observations to account for variations in spectral strength that planetary obliquity {(or observational phase)} can abiogenically induce when searching for biological influence in a signal across seasons, in order to avoid biosignature false-positives or false-negatives. Many repeated reflected light direct imaging observations of exoplanets are needed to constrain the effect obliquity has on multiple features across the recorded spectra to reduce this uncertainty. Repeated spectroscopic observations may also identify atmospheric features that are not detectable during other observations as the strength of both abiogenic and biogenic {gases} vary temporally \citep{olson_atmospheric_2018}. 

\subsection{Observational Capabilities} \label{subsec:error}
\begin{figure}[ht!]
\includegraphics[width=0.97\textwidth]{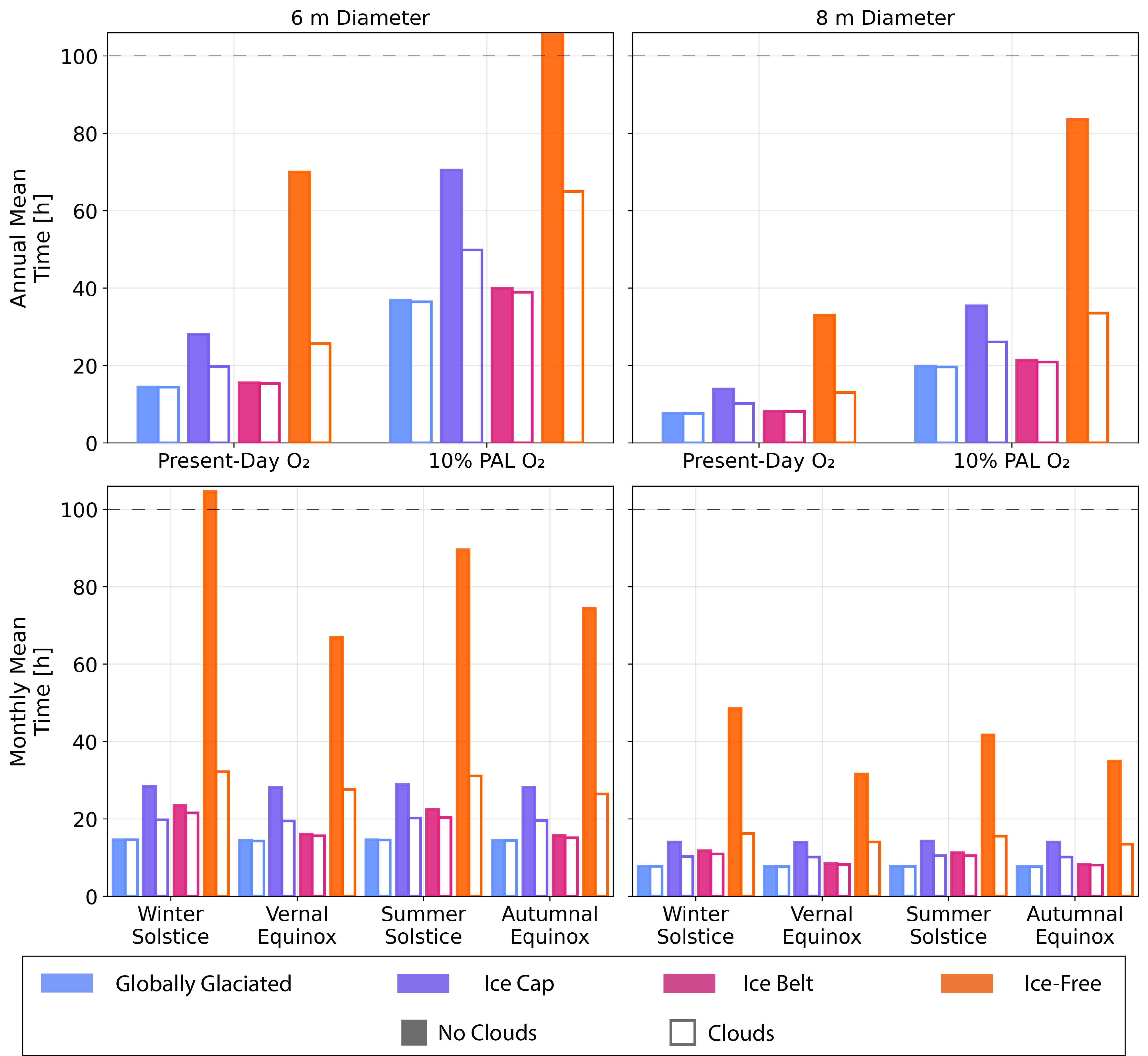}
\caption{Estimated minimum exposure times (in hours) required to achieve a $\geq$ 10 S/N observing the integrated O$_{2}$ band {(750-778 nm)} feature for planets as seen from 10 pc with a 6 m (1st column) and 8 m (2nd column) telescope diameter for all climate state experiments without and with clouds. We record the annual mean spectra exposure times for worlds with present-day and low O$_{2}$ in the top row, and the monthly mean spectra exposure times for worlds at each climate state at the solstices and equinoxes in the bottom row. Dashed line added to indicate 100 hour observation time cutoff.
\label{fig:ExposureTime}}
\end{figure}

We compute the estimated minimum exposure time in hours needed to detect an O$_{2}$ band feature {(calculated over the 750-778 nm bandpass)} at a S/N $\geq$ 10 on an exoplanet that is 10 pc away, as prescribed by HWO studies on noise \citep{feng_characterizing_2018, damiano_reflected_2022, damiano_reflected_2023, young_inferring_2024}, for all our planetary spectra experiments with both a 6 and 8 m diameter telescope mirror (Figure \ref{fig:ExposureTime}). The observations are {assumed to be at quadrature and} essentially photon noise dominated, which means that S/N is modulated by planetary albedo. 
{We select a photon noise dominated system as it is the most optimistic case in future direct imaging observations, and is the upper limit on how detectable a signal could be for a given instrument following past work \citep{stark_exoearth_2019, kofman_pale_2024}. HWO design is still changing which makes it challenging to explore telescope noise. Our work assuming a photon-noise dominated system leaves out complications and unknowns regarding the instrument specifics. While out of the scope of our project, this work motivates future telescope design trade studies and retrievals to investigate the impact of different potential direct imaging instrument design concepts on observational noise and feature detectability \citep{latouf_bayesian_2023, stark_paths_2024}. We thus ignore detector noise sources such as read noise and dark current, and treat telescope emission as negligible. Similarly, we ignore speckle and zodiacal noise, as they are currently difficult to predict and can degrade features, allowing us to isolate the intrinsic detectability of the spectroscopic signal \citep{kammerer_large_2022}. We explore how climate state influences spectra and feature detectability across simulations, treating these estimates as optimistic upper limits. We detail our instrument and model assumptions to calculate feature exposure time from S/N in Table \ref{SNTab}. We consider the same wavelength band definition and sampling assumptions when calculating exposure time as our equivalent width calculations.}

\begin{table}
\caption{{S/N Model Assumptions}}
\begin{center}
\begin{tabular}{l l}
\hline
Parameter & Values \\
\hline
Diameter (m) & 6, 8 \\
O$_{2}$ Bandpass (nm) & 750-778 \\
Resolving Power & R70 \\
No. of Detector Pixels & 10 \\
Contrast & 1e-10 \\
Temperature Optics T and I (K) & 273 \\
Throughput T and I (incl. QE) [AVG] & 0.28 \\
Emissivity T and I & 0.1 \\
\hline
\end{tabular}
\item{{We assume a photon noise dominated observation, with no other sources of noise, in exposure time calculations to determine feature exposure time. Parameters chosen from \cite{committee_for_a_decadal_survey_on_astronomy_and_astrophysics_2020_astro2020_pathways_2023} to investigate the impact of climate state on observations. The HWO design is currently only a recommendation and instrument configuration is subject to change.}}
\label{SNTab}
\end{center}
\end{table}

We find that the observation time required to confidently detect O$_{2}$ features varies considerably between our climate state experiments (both without and with clouds). We determine that the exposure time to detect the O$_{2}$ is considerably shorter with a 8 m telescope diameter than a 6 m telescope diameter across all experiments, which is particularly important for the design of future direct imaging telescopes. Not only do greater inscribed telescope diameters increase the yield of exo-Earth candidates that can be observed in HWO's survey \citep{kopparapu_exoplanet_2018, committee_for_a_decadal_survey_on_astronomy_and_astrophysics_2020_astro2020_pathways_2023}, we find that many observations with the larger 8 m diameter telescope require less than half the exposure time than the 6 m telescope would require to detect the same feature (Figure \ref{fig:ExposureTime}).
A telescope with a larger collecting area allows for O$_{2}$ features on the ice-free cloudless low oxygen world to be detectable in under 100 hours of exposure time, which would be otherwise undetectable with a 6 m diameter mirror. All other low oxygen experiments have required observation times under 40 hours with the 8 m diameter telescope, significantly improving the detectability of low O$_{2}$ exoplanets.
Because repeated spectroscopic observations will facilitate climate characterization and searching for biosignatures (Subsections \ref{subsec:season} and \ref{subsec:Biosigs}), reducing the required exposure time to confidently detect atmospheric features by increasing telescope diameter is highly desirable.

We also find that planetary climate state has significant impact on the observation time needed to resolve O$_{2}$ (55+ hour magnitude without clouds and 10+ hour difference with clouds for the present-day O$_{2}$ annual mean case), a time difference comparable to the impact of mirror size. Globally glaciated worlds have the lowest while the ice-free worlds have the greatest required exposure times to detect O$_{2}$ (for both annual and monthly mean spectra), as feature strength {and total signal strength} increase with planetary ice cover and albedo (Figures \ref{fig:HWO_EW} and \ref{fig:HWO_EW_Seasons}). 
The worlds with lower oxygen (10\% PAL) require longer exposure times {than} the planets with present-day O$_{2}$ at each climate state (Figure \ref{fig:ExposureTime}) as their atmospheric feature is smaller. 
Cloud cover increases albedo and thus has a significant impact reducing the exposure times for planets with low ice cover compared to their minimum observation times without clouds. For example, there is a $\sim$45 hour shorter exposure time for annual mean present-day O$_{2}$ ice-free world with clouds versus without. Thus, the feature detectability on a planet depends both on the climate state of the surface as well as on its cloud cover. 
There is temporal variation in the exposure time required to detect O$_{2}$ at different seasons (solstice vs. equinox) for the high obliquity (ice belt and ice-free) worlds in our monthly mean spectra (Figure \ref{fig:ExposureTime}). For high obliquity planets, the exposure time is longer during the winter and summer solstices than during the equinoxes. Lower obliquity worlds experience limited to no change in the required exposure time to resolve features at different times of observation. Measuring variations in required exposure time to detect {gases} like O$_{2}$ temporally (or a lack of variation), as well as the differences required exposure time between climates, may help inform the climate state of exoplanets in future spectroscopic observations.

\section{Conclusions} \label{sec:conclusions}

We demonstrate that it is possible to observationally distinguish between exoplanets at four end-member climate states that have the same atmospheric conditions, through resolvable differences in their {(at quadrature)} reflectance spectra. Icier worlds have greater apparent albedos with stronger absorption features, making observations more sensitive to their atmospheric signals, including potential biosignatures. There is therefore a tradeoff between the minimum necessary exposure time to detect atmospheric features on an exoplanet and its fractional habitability due to the lower reflectivity of ice-limited worlds. Clouds increase the apparent albedo spectra of low albedo worlds, {including their feature strengths. This} likely improves the detectability of {gases} on exoplanets with less ice {(when using the same wavelength band definition and sampling assumptions)} and {reduces the required} minimum exposure time compared to without clouds.
We find that worlds with high planetary obliquity vary in the brightness of their apparent albedo temporally. This creates potentially detectable abiogenic seasonality in high obliquity exoplanet spectra, where the strength of atmospheric features are time-varying across different seasons. 
As planetary radius is fixed, multiple spectroscopic observations that record temporal variability may therefore help to break radius-albedo degeneracy and inform the exoplanet climate state, especially when in combination with astrometry {to isolate the impact of phase angle on observations}. In future direct imaging missions using reflected light, repeated observations are needed to account for abiogenic temporal variation in spectral strength to assess planetary habitability and when searching for biosignatures.
 
\section{Acknowledgments} \label{sec:acknowledgments}
{We thank the anonymous reviewer for their thoughtful feedback that enhanced the manuscript significantly.} This project has been supported by NASA Exobiology for work on seasonality (grant No. 80NSSC20K1437), the NASA Habitable Worlds Program (grant No. 80NSSC20K1409), the NASA Interdisciplinary Consortia for Astrobiology Research (ICAR) program (grant no. 80NSSC21K0594), as well as the Heising-Simons Foundation (grant No. 2021-3127) to SLO.


\clearpage

\bibliography{gstar}{}
\bibliographystyle{aasjournal}

\end{document}